\documentclass[final, 3p, times, numbers, sort&compress]{elsarticle}

\usepackage{bm}
\usepackage{amsmath}
\usepackage{amsfonts}
\usepackage{amsthm}
\usepackage{amssymb}
\usepackage[utf8]{inputenc}
\usepackage{graphicx}
\usepackage[italicdiff]{physics}
\usepackage{wrapfig}
\usepackage{verbatim}
\usepackage{relsize}
\usepackage[breaklinks=true,colorlinks=true,linkcolor=blue,urlcolor=blue,citecolor=blue]{hyperref}

\usepackage{color}
\usepackage{xcolor}
\usepackage{soul}
\usepackage{subcaption}
\usepackage[utf8]{inputenc}

\begin{document}

\newcommand{\defeq}{\mathrel{\mathop:}=}

\title{A novel Bayesian approach to the computation\\of the configurational density of states}

\author[UNAB]{Felipe Moreno}
\ead{f.moreno.munoz@gmail.com}
\author[CCHEN,UNAB]{Sergio Davis}
\ead{sergio.davis@cchen.cl}
\author[UNAB]{Joaqu\'in Peralta}
\ead{joaquin.peralta@unab.cl}
\address[UNAB]{Departamento de F\'isica, Facultad de Ciencias Exactas, Universidad Andr\'es Bello, Sazi\'e 2212, Santiago, Chile.}
\address[CCHEN]{Research Center in the Intersection of Plasma Physics, Matter and Complexity (P$^2$mc), Comisi\'on Chilena de
Energ\'ia Nuclear, Casilla 188-D, Santiago 7600713, Chile.}
%\author{Felipe Moreno}
%\ead{f.moreno.munoz@gmail.com}
%\address{Departamento de F\'isica, Facultad de Ciencias Exactas, Universidad Andr\'es Bello, Sazi\'e 2212, Santiago, Chile.}
%\author{Sergio Davis}
%\ead{sergio.davis@cchen.cl}
%\address{Comisión Chilena de Energía Nuclear, Casilla 188-D, Santiago, Chile}
%\address{Departamento de F\'isica, Facultad de Ciencias Exactas, Universidad Andr\'es Bello, Sazi\'e 2212, Santiago, Chile.}
%\author{Joaqu\'in Peralta}
%\ead{joaquin.peralta@unab.cl}
%\address{Departamento de F\'isica, Facultad de Ciencias Exactas, Universidad Andr\'es Bello, Sazi\'e 2212, Santiago, Chile.}

\date{\today}

\begin{abstract}
In this work we develop and implement a novel Bayesian method for computing the DOS of a system. This method is based on the use of a test function with adjustable parameters and we use Bayes theorem to find the best parameters given a certain number of measurements done on the system. This measurements can be done in any ensemble defined by a distribution function. We found that the algorithm can find the DOS in a reasonable amount of time, and that if the test function is suitable enough, the DOS found by the algorithm is very close to the true DOS.
\end{abstract}
\maketitle

\section{Introduction}
The density of states (DOS), on its different flavors, such like energy, vibrational, configurational, among others~\cite{harrison1989}, is essentially the number of states per unit of energy. The configurational density of states, defined for a system with degrees of freedom $\bm x$ and potential energy $\Phi(\bm x)$ is given by
\begin{equation}
D(\phi) \defeq \int d\bm{x}\delta(\Phi(\bm x)-\phi).
\end{equation}

In physics and chemistry this DOS have an outstanding importance, because a large amount of information can be obtained from it. In spite of the enormous success in characterizing physical and chemical quantities based on the DOS~\cite{Gommes2012, Sarker2018, Garboczi1985, Mauro2007, YU2007, NAKAI2005}, it is not a easy to obtain function, which reduce its use, in spite of the their characteristics, on multiples research areas~\cite{Rose1996, Rathore2003, Rodgers1988}.

Different approaches has been implemented to find the DOS for a large variety of systems~\cite{methfessel1983, Do2013, ARIASOCA2019, Wang2001}. Nowadays one of the most successful
techniques correspond to determine the DOS is the one based in Wang Landau algorithm~\cite{Wang2001}, this technique is based on flat-histogram procedure and have been showed
that plenty of programming  techniques and improvements has been implemented during last decades~\cite{Dayal2004, Prellberg2004, liang_2006}. However it is not easy to calculate nor programming, and is even
more difficult to apply on specific scenarios, such like ab-initio.

While Wang-Landau has been very popular this time, there are different techniques for calculating DOS that are still in use, such as parallel tempering~\cite{Earl2005},
metadynamics~\cite{Micheletti2004,Laio2008,Barducci2008}, and other techniques based on flat histograms~\cite{Rathore2003,Rathore2003a}. In most of these cases, solution are focused on specific problems, and not always cover the variety of cases where DOS could be used. Motivated on this, and also in the incorporation of most general statistical techniques, we worked in a new methodology to determine the DOS.

In this work we present a novel procedure to obtain the energetic density of states (DOS) for solid materials, based on Bayesian method. Here we develop this novel technique by: i) incorporate the Bayesian Theorem on the procedure, ii) develop a modular implementation on a wide and popular programming language as Python3, and iii) the speed-up of certain function using C language.

In what follows we present the Algorithm description which connect the Bayesian theorem and the DOS (sec.~\ref{algorithm}). On section ~\ref{implementation} a brief
description of the implementation of the algorithm is presented using Python programming language. Section~\ref{results} present cases of study for different scenarios,
such like classical solid structure using a Lennard-Jones interatomic potential, and the Ising model. Finally on section~\ref{conclusions} conclusions and scope of our
work is discussed.

\section{Algorithm}
\label{algorithm}

\noindent
Suppose the DOS of a system is completely determined by $m$ parameters $\boldsymbol\theta=(\theta_1,\ldots,\theta_m)$
\begin{equation}
D(\phi)=D(\phi;\boldsymbol\theta).
\end{equation}

The probability that an ensemble $S$ has a certain energy $\phi$ is
\begin{equation}
P(\phi;S|\boldsymbol\theta)=\frac{\rho(\phi;S)D(\phi;\boldsymbol\theta)}{Z(\boldsymbol\theta;S)}
\end{equation}
where $\rho(\phi;S)$ is the distribution that defines the ensemble $S$ and $Z(\boldsymbol\theta;S)$ is the partition function defined by
\begin{equation}
Z(\boldsymbol\theta;S)=\int d\phi\, \rho(\phi;S)D(\phi;\boldsymbol\theta).
\end{equation}

Suppose we make $n$ measurements of the ensemble energy, obtaining $\phi_1,\ldots,\phi_n$. We want to know what is the most probable values of the parameters $\boldsymbol\theta$ compatible with those measurements. We accomplish this by computing the maximum of $P(\boldsymbol\theta|\phi_1,\ldots,\phi_n;S)$. In practice, it is better to compute the maximum of the
logarithm of $P(\boldsymbol\theta|\phi_1,\ldots,\phi_n;S)$, that is,
\begin{equation}
\label{eq:maximize}
\pdv{\theta_k}\ln P(\boldsymbol\theta|\phi_1,\ldots\phi_n;S)=0\qquad\qquad k=1,\ldots,m.
\end{equation}

\noindent
Because the measurements are considered to be statistically independent, we have
\begin{equation}
\label{eq:likelihood}
P(\phi_1,\ldots,\phi_n|\boldsymbol{\theta},S) = \prod_{i=1}^n \frac{\rho(\phi_i;S)D(\phi_i; \boldsymbol{\theta})}{Z(\boldsymbol{\theta}; S)}
\end{equation}
and, by using Bayes theorem~\cite{Sivia2006} we have
\begin{equation}
\label{eq:bayes}
\begin{split}
P(\boldsymbol\theta|\phi_1,\ldots\phi_n;S) & = \frac{P(\boldsymbol\theta|I_0)P(\phi_1,\ldots,\phi_n|\boldsymbol{\theta},S)}{P(\phi_1,\ldots,\phi_n;S)} \\
& = \frac{P(\boldsymbol\theta|I_0)}{P(\phi_1,\ldots,\phi_n;S)}\prod_{i=1}^n\frac{\rho(\phi_i;S)D(\phi_i;\boldsymbol\theta)}{Z(\boldsymbol\theta;S)} \\
& =\frac{1}{\eta}P(\boldsymbol\theta|I_0)\frac{\prod_{i=1}^nD(\phi_i;\boldsymbol\theta)}{Z(\boldsymbol\theta;S)^n}
\end{split}
\end{equation}
where
\begin{equation}
\eta \defeq \frac{\prod_{i=1}^n\rho(\phi_i;S)}{P(\phi_1,\ldots,\phi_n|I_0)}
\end{equation}
is an irrelevant normalization constant that does not depend on $\boldsymbol\theta$.

\noindent
In this way, Eq.~\ref{eq:maximize} can be written as
%\vspace{1cm}
%
%\begin{widetext}
\begin{equation}
\label{eq:wide}
\pdv{\theta_k}\ln P(\boldsymbol\theta|\phi_1,\ldots\phi_n;S) = \pdv{\theta_k}\ln P(\boldsymbol\theta|I_0)+\sum_{i=1}^n\pdv{\theta_k}\ln D(\phi_i;\boldsymbol\theta)
  -n\pdv{\theta_k}\ln Z(\boldsymbol\theta;S) = 0.
\end{equation}
%\end{widetext}

\noindent
In the limit $n\rightarrow \infty$ the first term on the right-hand side of Eq.~\ref{eq:wide} is negligible, and Eq.~\ref{eq:maximize} becomes
\begin{equation}
\label{eq:ln_Z}
\pdv{\theta_k}\ln Z(\boldsymbol\theta;S)=\frac 1n\sum_{i=1}^n\pdv{\theta_k}\ln D(\phi_i;\boldsymbol\theta)
\end{equation}
for $k=1,\ldots,m$. Notice that the right-hand side is the average of a known function of the measurements, while the left-hand side is given by
\begin{equation}
\begin{split}
\pdv{\theta_k}\ln Z(\boldsymbol\theta;S) & = \frac{1}{Z(\boldsymbol\theta;S)}\int d\phi\,\rho(\phi;S)\pdv{\theta_k}D(\phi;\boldsymbol\theta) \\
 & = \frac{1}{Z(\boldsymbol\theta)}\int d\phi\,\rho(\phi;S)D(\phi;\boldsymbol\theta)\pdv{\theta_k}\ln D(\phi;\boldsymbol\theta) \\
 & = \ev{\pdv{\theta_k}\ln D(\phi;\boldsymbol\theta)}_S,
\end{split}
\end{equation}
that is, the expectation value over the ensemble of the same function that is averaged on the right-hand side. In other words, we have that the condition
that fixes $\boldsymbol\theta$ is simply
\begin{equation}
\label{eq:1ensemble}
\ev{\pdv{\theta_k}\ln D(\phi;\boldsymbol\theta)}_S=\overline{\pdv{\theta_k}\ln D(\phi;\boldsymbol\theta)}
\end{equation}
for $k=1,\ldots,m$. This condition can be understood as the validity of the law of large numbers, namely the identification of the expected value with the data average, for the logarithmic derivatives of the density of states.

A straightforward generalization is to consider $r$ groups of $n_1,\ldots,n_r$ measurements obtained from different ensembles $S_1,\ldots,S_r$
respectively. Then  Eq.~\ref{eq:bayes} generalizes to
\begin{equation}
P(\boldsymbol\theta|\phi_1,\ldots\phi_n;S_1,\ldots,S_r)=\frac{1}{\eta}P(\boldsymbol\theta)\frac{\prod_{i=1}^N D(\phi_i;\boldsymbol\theta)}{\prod_{j=1}^r[Z(\boldsymbol\theta;S_j)]^{n_j}}
\end{equation}
where $N=\sum_{j=1}^rn_j$ is the total number of measurements, and Eq.~\ref{eq:1ensemble} generalizes to
\begin{equation}
\label{eq:THE_equation}
\sum_{j=1}^r\frac{n_j}{N}\ev{\pdv{\theta_k}\ln D(\phi;\boldsymbol\theta)}_{S_j}=\overline{\pdv{\theta_k}\ln D(\phi;\boldsymbol\theta)}
\end{equation}
for $k=1,\ldots,m$ where the average in the right hand side is performed over all $N$ measurements. This algorithm is based on the solution of Eq.~\ref{eq:THE_equation} by
numerical methods. An equivalent problem is to search for the global minimum of
\begin{equation}
\label{eq:epsilon2}
\epsilon^2(\boldsymbol\theta)=\sum_{k=1}^m\left[\sum_{j=1}^r\frac{n_j}{N}\ev{\pdv{\theta_k}\ln D(\phi;\boldsymbol\theta)}_{S_j}
\hspace{-10pt}-\overline{\pdv{\theta_k}\ln D(\phi;\boldsymbol\theta)}\right]^2
\end{equation}
by using numerical minimization methods such as particle swarm optimization (PSO)~\cite{Olsson2011}.
\\\\
Now we can formulate a practical method of computing the DOS, using Eq.~\ref{eq:epsilon2}. Firstly, we select a set of canonical ensembles $S_1,\ldots,S_r$ such that \[\rho(\phi;S_j)\propto \exp(-\beta_j\phi)\] and we take $n_j$ samples of energy respectively. The ensemble information ($\beta_j$) and the measurements ($\phi_i$) are stored. Secondly, we choose a suitable function $\ln D(\phi;\boldsymbol\theta)$ and then we run a minimization algorithm on the sampled data in order to find the global minimum of $\epsilon^2(\boldsymbol\theta)$. Once the global minimum has been found, the parameters $\boldsymbol\theta$ that minimize $\epsilon^2(\boldsymbol\theta)$ are the ones that maximize the probability that $D(\phi;\boldsymbol\theta)$ is the true DOS, within the chosen parametrization. If the function chosen is suitable enough, we expect this function to be very close to the true DOS.

Alternatively, we could have chosen a different set of ensembles, for instance, the microcanonical configurational ensemble \[\rho(\phi; S_j) \propto (E_j-\phi)^{\frac{d\mathcal{N}}{2}-1}\] for different total energies $E_1,\ldots,E_r$ where $d$ is the number of spatial dimensions.

\section{Implementation}
\label{implementation}

\noindent
The algorithm consists of two parts. The first one is to generate measurements of the ensembles energy. We do this by using a canonical Monte Carlo Metropolis algorithm.
We set an initial state $\vb x=(x_1,\ldots,x_{\mathcal{N}})$ and we propose a transition to a new state $\vb x'$ which is accepted with probability
\begin{equation}
P(\vb x\rightarrow \vb x'|\beta)=\min\left(1,e^{-\beta(\Phi(\vb x')-\Phi(\vb x))}\right)
\end{equation}
where $\Phi(\vb x)$ is the configurational energy of state $\vb x$. We vary $\beta$ in order to produce data for different ensembles. We take samples of the ensemble energy
after a certain number of calibration steps have been performed following each variation of $\beta$. This part of the algorithm will generate two files, one with the set of
temperatures $\beta_j$ that define each ensemble, and one with the sampled energies.
\\\\
The second part of the algorithm is to run a PSO algorithm on the sampled data in order to find the global minimum of $\epsilon^2(\boldsymbol\theta)$. Since the computation of $\epsilon^2$ is the computationally expensive part of the algorithm, it has been coded in C++ and loaded as a library into a easier to handle Python code.
\\\\
Furthermore, we can speed up the minimization by considering a very common special case. Suppose that $\ln D(\phi,\boldsymbol\theta)$ depends linearly
on the parameters $\boldsymbol\theta$, that is
\begin{equation}
    \ln D(\phi;\boldsymbol\theta) = \sum_{k=1}^m\theta_kf_k(\phi)
\end{equation}
where we can interpret the set of functions $\{f_1(\phi),\ldots,f_m(\phi)\}$ as a basis for the representation of the density of states. Inserting this into
Eq.~\ref{eq:epsilon2} we obtain
\begin{equation}
\epsilon^2(\boldsymbol\theta)=\sum_{k=1}^m\left\{\sum_{j=1}^r\frac{n_j}{N}\ev{f_k(\phi)}_{S_j}-\overline{f_k(\phi)}\right\}^2,
\end{equation}
where the terms $\overline{f_k(\phi)}$ do not depend on $\boldsymbol\theta$ thus they will only need to be computed once. Since these are the only terms that depend on the sampled data, we can make the minimization time independent of the size of this data.

%The estimation of the statistical error of the DOS reconstruction is not a trivial task\fixme{[CITES]. However, in this case we propose the use of the Bayesian information criterion (BIC)}~\cite{Schwarz1978}, defined by

To estimate the statistical error of the DOS reconstruction, we peopose the use of the Bayesian information criteria (BIC)~\cite{Schwarz1978}, which is defined by:

\begin{equation}
\text{BIC} \defeq -2\ln P(\phi_1, \ldots, \phi_N|\boldsymbol \theta) + m\ln N
\end{equation}
and such that we should prefer the model with the lowest BIC to reduce the error. By using Eq.~\ref{eq:likelihood} we have
\begin{equation}
\text{BIC} = -2N\Big(\overline{\ln \rho(\phi;S)}+\overline{\ln D(\phi;\boldsymbol{\theta})} - \ln Z(\boldsymbol{\theta}; S)\Big) + m\ln N.
\end{equation}

Because the first term inside the parenthesis does not depend on the parameterization $\boldsymbol{\theta}$ we can instead compare two models $D_1(\phi; \bm{\theta}_1)$ and $D_2(\phi; \boldsymbol{\theta}_2)$ for the same set of ensembles and same data through the quantity
\begin{equation}
\text{BIC}^* = -2\Big(N\overline{\ln D(\phi;\boldsymbol{\theta})} - \sum_{j=1}^r n_j\ln Z(\boldsymbol{\theta}; S_j)\Big) + m\ln N.
\end{equation}

\section{Results}
\label{results}
\noindent

The algorithm and implementation have been tested using a variety of systems in order to evaluate the accuracy of the suggested methodology.

\subsection{Harmonic Oscillator}
\noindent
We consider a system of $\mathcal{N}=100$ identical harmonic oscillators,
\begin{equation}
\Phi(\vb x)=\frac 12m\omega^2\sum_{i=1}^{\mathcal{N}}x_i^2.
\end{equation}
Instead of $\beta$ we have used $k_BT=\beta^{-1}$ in the range $k_BT\in [0.05,5]$ at steps $\Delta (k_BT) = 0.05$. We consider the energy of the last $1\,000$ steps of simulation---in  a total of 10\,000---for each $k_BT$. That is, we produce $r=100$ values of $k_BT$ and $N=100\,000$ values of energy $\phi$.
\\\\
Next, we chose the following test function with one parameter ($m=1$),
\begin{equation}
\ln D(\phi;\boldsymbol\theta)=\theta_1\ln\phi.
\end{equation}

\begin{figure}[h!]
\begin{center}
\includegraphics[width=8cm]{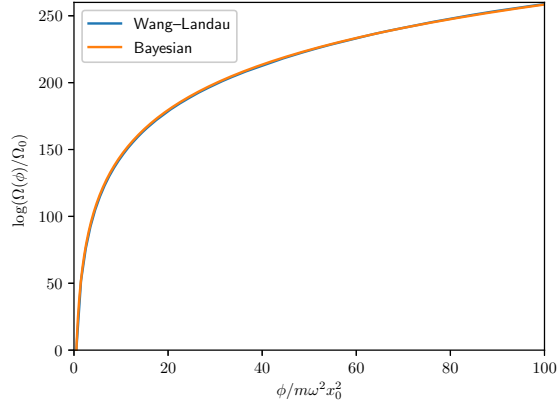}
\end{center}
\caption{Configurational DOS for a system of $\mathcal{N}$=100 classical harmonic oscillators.}
\label{fig:ho}
\end{figure}

We run the PSO algorithm on this data using 1760 particles and after 37 iterations we found that $\theta_1=49.001$, which closely matches the analytical result
\[D(\phi)\propto\phi^{\mathcal{N}/2-1}.\]

The results for this case are presented on figure~\ref{fig:ho}, we observe an excellent agreement with the WL algorithm result.

\subsection{Ising model}
The Ising model is a well-known model of spins. The model consists in a lattice filled with spins that can take values $\sigma=\pm 1$. The model energy is computed according to
\begin{equation}
    \phi = -J\sum_{\langle i,j\rangle}\delta(\sigma_i, \sigma_j).
\end{equation}
where the sum is over adjacent pairs and $\delta(\sigma_i, \sigma_j)$ is 1 if $\sigma_i=\sigma_j$ and 0 otherwise. We use an Ising model on a $10\times 10$ lattice and we approximate the DOS using two 1-parameter test functions. Since the DOS is defined up to a constant, we require that the test functions vanish at $\phi=\pm 200 J$ which are the energies less populated. The test functions will be
\begin{align}
    \log D_1(\phi;\boldsymbol\theta) &= \theta_1\left[\left(\frac{\phi}{200}\right)^2-1\right]\\
    \log D_2(\phi;\boldsymbol\theta) &=
    \theta_1\cos\left(\frac{\phi}{2\cdot200}\right)
\end{align}

\begin{figure}[h!]
\begin{center}
\includegraphics[width=8cm]{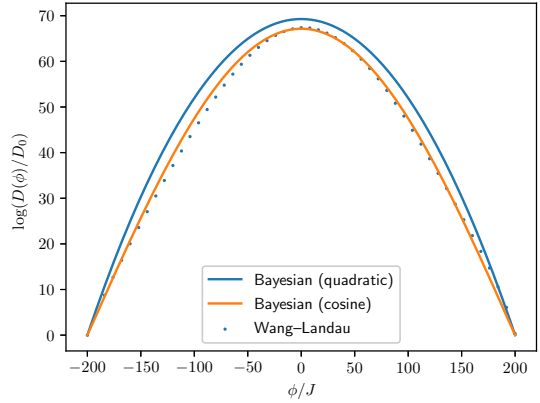}
\end{center}
\caption{Configurational DOS for an Ising system of 100 spins on a $10 \times 10$ lattice using a quadratic and a cosine approximation for $\ln D(\phi)$.}
\label{fig:ising}
\end{figure}

We found the parameters $\theta_1=-69.2799$ for the first function and $\theta_1=67.1554$ for the second one. The BIC is smaller for the cosine function $3.40277\times 10^7 < 3.41584\times 10^7$. The results for this case are presented on figure~\ref{fig:ising}, we observe a better agreement with WL algorithm on the function with the lower BIC value.

\subsection{Lennard-Jones}
\begin{figure}[h!]
\begin{center}
\includegraphics[width=8cm]{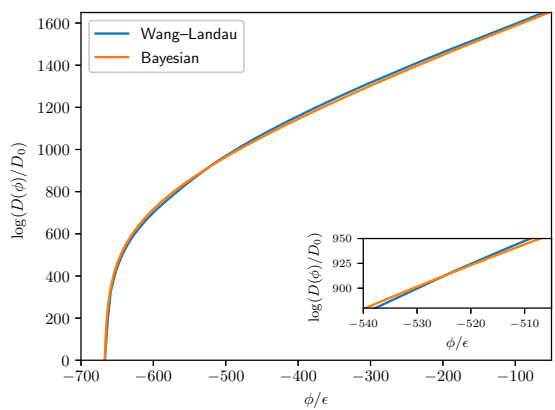}
\end{center}
\caption{Configurational DOS for a Lennard-Jones system of 120 particles assuming $\log D(\phi)$ is of the form shown in Eq.~\ref{eq:log_D_lj}.}
\label{fig:lj}
\end{figure}

Finally, we study a system that needs 2 parameters to get a good approximation. A Lennard--Jones gas is a gas of particles whose interaction energy is
\begin{equation}
    \phi = \sum_{i,j}4\varepsilon\left[\left(\frac{\sigma}{r_{ij}}\right)^6-\left(\frac{\sigma}{r_{ij}}\right)\right],
\end{equation}
where the sum is over every pair. For this system we consider the addition of a logarithm and a linear function.
\begin{equation}\label{eq:log_D_lj}
    \log D(\phi;\boldsymbol\theta)=\theta_1\log (\phi-\phi_0)+\theta_2\phi
\end{equation}
where $\phi_0$ is the minimum energy.
\\\\
We found $\theta_1=1.52994\times 10^2$ and $\theta_2=1.09103$. The drawback here is that this model is unable to produce the phase transition. In spite of this, as we can observe on figure~\ref{fig:lj}, there is a clear insight in the figure whit a results that behave different on the two main branches of the WL algorithm. For energies lower than -530 , the Bayesian estimation is above the WL approximation. On the other side, with energies above the -530 the Bayesian estimation is below the WL results. This suggest that our model it could be splitted on different branches when we know beforehand about the presence of a phase transition.

\section{Conclusions}
\label{conclusions}
\noindent
We have presented a new technique to determine the Density of States. Different simulation examples have been used: Ising model, Lenard-Jones, and classical harmonic oscillator. Our results shows a good agreement with the known theoretical results. Moreover this methodology shows an improvement on computational efforts to be required to determine the DOS. The requirement of an input function, its an strength when some insights are known beforehand, otherwise a large functional base can always be used. If the functional form is improved the BIC error procedure will return the most effective evaluation of the DOS.

\nocite{Habeck2007}   % DOS via Bayes theorem
\nocite{Tkacik2014}   % Wang-Landau used in inference

\nocite{Moreno2018}   % Potts model
\nocite{Farias2021}   % Potts model off-lattice
\nocite{Davis2020d}   % Coulomb particles inside a container

\section{Acknowledgments}
FM thanks to Universidad Andrés Bello for providing the doctorate scholarship and acknowledges funding from ANID BECAS/DOCTORADO NACIONAL 21212267. SD acknowledges partial funding from ANID PIA ACT172101 grant. Computational work was supported by the supercomputing infrastructures of the NLHPC (ECM-02), and FENIX (UNAB).

\bibliography{dos}

\providecommand{\noopsort}[1]{}\providecommand{\singleletter}[1]{#1}%
\begin{thebibliography}{10}

\bibitem{harrison1989}
W.A. Harrison.
\newblock {\em Electronic Structure and the Properties of Solids: The Physics
  of the Chemical Bond}.
\newblock Dover Books on Physics. Dover Publications, 1989.

\bibitem{Gommes2012}
C.~J. Gommes, Y.~Jiao, and S.~Torquato.
\newblock Density of states for a specified correlation function and the energy
  landscape.
\newblock {\em Phys. Rev. Lett.}, 108:080601, Feb 2012.

\bibitem{Sarker2018}
P.~Sarker, T.~Harrington, C.~Toher, C.~Oses, M.~Samiee, J.-P. Maria, D.~W.
  Brenner, K.~S. Vecchio, and S.~Curtarolo.
\newblock High-entropy high-hardness metal carbides discovered by entropy
  descriptors.
\newblock {\em Nature {C}ommunications}, 9:1--10, 2018.

\bibitem{Garboczi1985}
E.~J. Garboczi and M.~F. Thorpe.
\newblock Density of states for random-central-force elastic networks.
\newblock {\em Phys. Rev. B}, 32:4513--4518, Oct 1985.

\bibitem{Mauro2007}
John~C. Mauro, Roger~J. Loucks, Jitendra Balakrishnan, and Srikanth Raghavan.
\newblock Monte carlo method for computing density of states and quench
  probability of potential energy and enthalpy landscapes.
\newblock {\em The Journal of Chemical Physics}, 126(19):194103, 2007.

\bibitem{YU2007}
S.-Q. Yu, J.-B. Wang, D.~Ding, S.R. Johnson, D.~Vasileska, and Y.-H. Zhang.
\newblock Impact of electronic density of states on electroluminescence
  refrigeration.
\newblock {\em Solid-State Electronics}, 51(10):1387--1390, 2007.
\newblock Special Issue: Papers Selected from the NGC2007 Conference.

\bibitem{NAKAI2005}
N.~Nakai, P.~Miranović, M.~Ichioka, and K.~Machida.
\newblock Theoretical study on the field dependence of the zero energy density
  of states in an anisotropic gap superconductors.
\newblock {\em Journal of Physics and Chemistry of Solids}, 66(8):1362--1364,
  2005.
\newblock Proceedings of the ISSP International Symposium (ISSP-9)on Quantum
  Condensed System.

\bibitem{Rose1996}
Helge Ros{\'e}, Werner Ebeling, and Torsten Asselmeyer.
\newblock The density of states --- a measure of the difficulty of optimisation
  problems.
\newblock In Hans-Michael Voigt, Werner Ebeling, Ingo Rechenberg, and Hans-Paul
  Schwefel, editors, {\em Parallel Problem Solving from Nature --- PPSN IV},
  pages 208--217, Berlin, Heidelberg, 1996. Springer Berlin Heidelberg.

\bibitem{Rathore2003}
N.~Rathore, T.~A.~Knotts IV, and J.~J. de~Pablo.
\newblock Configurational temperature density of states simulations of
  proteins.
\newblock {\em Biophys. J.}, 85:3963--3968, 2003.

\bibitem{Rodgers1988}
G.~J. Rodgers and A.~J. Bray.
\newblock Density of states of a sparse random matrix.
\newblock {\em Phys. Rev. B}, 37:3557--3562, Mar 1988.

\bibitem{methfessel1983}
M~S Methfessel, M~H Boon, and F~M Mueller.
\newblock Analytic-quadratic method of calculating the density of states.
\newblock {\em Journal of Physics C: Solid State Physics}, 16:L949--L954, 09
  1983.

\bibitem{Do2013}
Hainam Do and Richard~J. Wheatley.
\newblock Density of states partitioning method for calculating the free energy
  of solids.
\newblock {\em Journal of Chemical Theory and Computation}, 9(1):165--171,
  2013.
\newblock PMID: 26589019.

\bibitem{ARIASOCA2019}
Thomas~Aquino Ariasoca, Sholihun, and Iman Santoso.
\newblock Trotter-suzuki-time propagation method for calculating the density of
  states of disordered graphene.
\newblock {\em Computational Materials Science}, 156:434--440, 2019.

\bibitem{Wang2001}
F.~Wang and D.~P. Landau.
\newblock Efficient, multiple-range random walk algorithm to calculate the
  density of states.
\newblock {\em Phys. Rev. Lett.}, 86:2050--2053, 2001.

\bibitem{Dayal2004}
P.~Dayal, S.~Trebst, S.~Wessel, D.~W\"urtz, M.~Troyer, S.~Sabhapandit, and
  S.~N. Coppersmith.
\newblock Performance limitations of flat-histogram methods.
\newblock {\em Phys. Rev. Lett.}, 92:097201, Mar 2004.

\bibitem{Prellberg2004}
Thomas Prellberg and Jaros\l{}aw Krawczyk.
\newblock Flat histogram version of the pruned and enriched rosenbluth method.
\newblock {\em Phys. Rev. Lett.}, 92:120602, Mar 2004.

\bibitem{liang_2006}
Faming Liang.
\newblock A theory on flat histogram monte carlo algorithms.
\newblock {\em Journal of Statistical Physics}, 122:511--529, 01 2006.

\bibitem{Earl2005}
D.~J. Earl and M.~W. Deem.
\newblock Parallel tempering: Theory, applications, and new perspectives.
\newblock {\em Phys. Chem. Chem. Phys.}, 7:3910--3916, 2005.

\bibitem{Micheletti2004}
C.~Micheletti, A.~Laio, and M.~Parrinello.
\newblock Reconstructing the density of states by history-dependent
  metadynamics.
\newblock {\em Phys. Rev. Lett.}, 92:170601, 2004.

\bibitem{Laio2008}
A.~Laio and F.~L. Gervasio.
\newblock Metadynamics: a method to simulate rare events and reconstruct the
  free energy in biophysics, chemistry and materials science.
\newblock {\em Rep. Prog. Phys.}, 71:126601, 2008.

\bibitem{Barducci2008}
A.~Barducci, G.~Bussi, and M.~Parrinello.
\newblock Well-tempered metadynamics: A smoothly converging and tunable
  free-energy method.
\newblock {\em Phys. Rev. Lett.}, 100:020603, 2008.

\bibitem{Rathore2003a}
N.~Rathore, T.~A. Knotts, and J.~J. de~Pablo.
\newblock Density of states simulations of proteins.
\newblock {\em J. Chem. Phys.}, 118(9):4285--4290, 2003.

\bibitem{Sivia2006}
D.~S. Sivia and J.~Skilling.
\newblock {\em Data Analysis: A Bayesian Tutorial}.
\newblock Oxford University Press, 2006.

\bibitem{Olsson2011}
A.~E. Olsson.
\newblock {\em Particle {S}warm {O}ptimization: {T}heory, techniques and
  applications}.
\newblock Nova Science Publishers, 2011.

\bibitem{Schwarz1978}
G.~E. Schwarz.
\newblock Estimating the dimension of a model.
\newblock {\em Ann. Statist.}, 6:461--464, 1978.

\bibitem{Habeck2007}
M.~Habeck.
\newblock Bayesian reconstruction of the density of states.
\newblock {\em Phys. Rev. Lett.}, 98:200601, 2007.

\bibitem{Moreno2018}
F.~Moreno, S.~Davis, C.~Loyola, and J.~Peralta.
\newblock Ordered metastable states in the {P}otts model and their connection
  with the superheated solid state.
\newblock {\em Phys. A}, 509:361--368, 2018.

\bibitem{Farias2021}
C.~Farías and S.~Davis.
\newblock {Multiple metastable states in an off-lattice {P}otts model}.
\newblock {\em Phys. A}, 581:126215, 2021.

\bibitem{Davis2020d}
S.~Davis, J.~Jain, and B.~Bora.
\newblock Computational statistical mechanics of a confined, three-dimensional
  {C}oulomb gas.
\newblock {\em Phys. Rev. E}, 102:042137, 2020.

\end{thebibliography}
\bibliographystyle{unsrt}

\end{document}